\documentclass{IEEEtran}
\usepackage[dvipsnames]{xcolor}
\usepackage{cite}
\usepackage{amsmath,amssymb,amsfonts}
\usepackage{mathtools}
\usepackage{lipsum}
\usepackage{algorithmic}
\usepackage{graphicx}
\usepackage{textcomp}
\usepackage{booktabs}
\usepackage{esint}
\def\BibTeX{{\rm B\kern-.05em{\sc i\kern-.025em b}\kern-.08em
    T\kern-.1667em\lower.7ex\hbox{E}\kern-.125emX}}
\begin{document}
\title{Hollow Rectangular Waveguide-fed Holographic Beamforming Antenna Additively Manufactured (3D Printed) with Conductive Polymer}
\author{Insang Yoo, \IEEEmembership{Member, IEEE}, Jonah Gollub, \IEEEmembership{Member, IEEE}, Shengrong Ye, Allen Gray, Okan Yurduseven \IEEEmembership{Senior Member, IEEE}, Manohar D. Deshpande \IEEEmembership{Senior Member, IEEE} and David R. Smith \IEEEmembership{Senior Member, IEEE}
\thanks{I. Yoo, J. Gollub and D. R. Smith are with the Department of Electrical and Computer Engineering, Duke University, Durham, NC, 27708 USA (e-mail: insangyoo1@gmail.com). S. Ye and A. Gray are with Multi3D Inc., Middlesex, NC, 27557 USA. O. Yurduseven is with the School of Electronics, Electrical Engineering and Computer Science, Queen's University Belfast, Belfast, UK. M. Deshpande is with the Microwave Instrument Technology Branch (Code 555), NASA Goddard Space Flight Center, Greenbelt, MD 20771 USA.}}

\maketitle

\begin{abstract}
We present the design and fabrication of 3D printed holographic beamforming antennas. The antennas utilize additively manufactured hollow rectangular waveguides that feed radiating rectilinear slots inserted into the upper conducting wall. The lengths of the individual slots are altered to implement a holographic beamforming solution designed using a coupled dipole formalism. For rapid verification, the designed antennas are fabricated using a desktop dual-extrusion fused filament 3D printer. The body of each antenna and its inner conducting surface are respectively printed using polylactic acid and biodegradable conductive polyester composite material (i.e., Electrifi), which is later deposited with a layer of copper on its surface to improve surface conductivity and reduce surface roughness. The beamforming performance of the fabricated antennas is confirmed via experiments. The 3D printed metasurface antennas using the proposed fabrication technique illustrate emerging capabilities in the rapid prototyping of complex electromagnetic structures.
\end{abstract}


\section{Introduction} \label{sec:introduction}

\IEEEPARstart{H}{}olographic beamforming antennas\textemdash initially suggested by \cite{oliner1959guided}\textemdash are of considerable interest for their abilities to generate single or multiple directive beams from a series of artificially structured scattering elements patterned into waveguiding structures \cite{sievenpiper2005holographic,fong2010scalar,Minatti2011,patel2011printed,gonzalez2017multibeam,yurduseven2017dual,smith2017analysis,boyarsky2021electronically}. In these devices, the scattering elements are designed to interact with a feed wave and implement a desired current distribution over an aperture, resulting in constructive interference of the radiated fields in desired directions. The devices can implement this beamforming or beam-steering capability using low-power switchable elements (e.g., diodes, liquid crystals) \cite{boyarsky2021electronically,stevenson2016metamaterial} avoiding costly, power-hungry RF components (e.g., phase shifters) associated with conventional phased arrays and other electronically scanned antennas (ESAs).

Among the plethora of possible implementations of such devices, waveguide or cavity-backed configurations\textemdash which we refer to as waveguide-backed metasurfaces\textemdash have proven advantageous in their pattern synthesis capabilities \cite{smith2017analysis} and have achieved significant commercialization over the past several years \cite{stevenson2016metamaterial,stevenson2018high,devadithya2017gpu,staff2019holographic}. In such antenna configurations, electrically small apertures, or elements, are inserted into one of the conducting surfaces of a waveguide, from which the guided mode field radiates. A complex weight is introduced by each element relating to its geometry and electrical properties. In an alternative description, each sub-wavelength element can be described as a polarizable electric or magnetic dipole, with its polarizability playing a role analogous to the weighting factor in an array antenna element. Thus, a desired radiation pattern can be formed by backpropagating the desired fields to the aperture plane, then selecting the weights---or polarizabilities---of the elements such that the desired aperture fields are well-approximated by the feed wave scattered by the dipole-like elements. \cite{smith2017analysis,yurduseven2017dual,johnson2015sidelobe,boyarsky2021electronically}.


Waveguide-backed metasurface antennas with holographic beamforming capability are now routinely fabricated using standard printed circuit board (PCB) processing techniques \cite{smith2017analysis,yurduseven2017dual,johnson2015sidelobe,boyarsky2021electronically}. These techniques are well-established, providing high accuracy and fidelity in fabrication; however, they often suffer from inherent dielectric losses associated with wave propagation in the substrate, which becomes even more severe at higher frequencies. Also, they are often limited to planar structures, requiring additional processes for fabricating three-dimensional (3D) structures. As a promising alternative, additive manufacturing\textemdash referred to as 3D printing (3DP)\textemdash has recently drawn much interest in the RF community as the 3DP technique offers new design freedom in building 3D RF systems, including lenses and antennas with complex shapes \cite{zhang2016design,wu20193,tak20183,yurduseven2017computational,le20163d,jun20183}. More recently, a 3D printable, highly conductive filament has introduced a novel way to metalize 3D printed dielectric structures, simplifying the fabrication of 3D printed electromagnetic designs, such as horn antennas \cite{kim2019one}, metamaterials \cite{xie2017microwave}, cavity-backed metasurface antennas \cite{yurduseven2017computational}, and near-field focusing structures \cite{yurduseven20193d}. The 3DP technique has emerged as a promising technology to enable novel electromagnetic designs not amenable to traditional planar fabrication.

To explore the application of 3D printing for metasurface antennas, we present here the design and fabrication of 3D printed, passive holographic beamforming antennas. For the design, we use rectilinear slots as radiating elements, expressing the transmitted magnitude from each slot as an effective complex polarizability. The polarizability properties of a given slot can be tuned by adjusting the slot length. The polarizability distribution is then determined using holographic principles to form a directed beam in a specific direction. Fabrication and metalization of the designed antennas are performed using a single-step 3D printing process, with an additional step of depositing a thin copper layer to improve the surface conductivity. The use of conductive polymer as a seed layer for electroplating can avoid complex processes for metalization. The operation of the designed beamforming antennas is confirmed via experiments on fabricated samples. The design and fabrication process outlined in this work paves the way towards rapid prototyping of complex electromagnetic structures, including planar or curved metasurface antennas.

This work builds on prior reports of 3D printed slotted waveguide antennas, such as those in \cite{mckerricher2015lightweight,le20163d,tak20183,zhao2019fully}. Distinct from these previous examples, here we make use of subwavelength-spaced radiating slots, homogenizing the aperture field on the individual slots to arrive at an effective magnetic polarizability. The dipolar modeling approach allows each slot to be treated as an equivalent current, thus simplifying the analysis and design of the antennas \cite{pulido2017polarizability}. The dipole model contrasts, for example, with Elliott's design procedure \cite{elliot2006antenna}, where the individual slots are modeled using the aperture field without homogenization. In terms of fabrication, the metallization of the antennas is achieved by selective electroplating of the conductive polymer composite \cite{kim2019one}, which does not require complex processes such as electroless nickel plating \cite{le20163d,zhao2019fully}, spray painting \cite{tak20183}, or doctor blade coating with a conductive silver ink \cite{mckerricher2015lightweight}.


\section{Design of Holographic Beamforming Antennas } \label{sec:theory}

\begin{figure}[t]
\centerline{\includegraphics[width=3.4in]{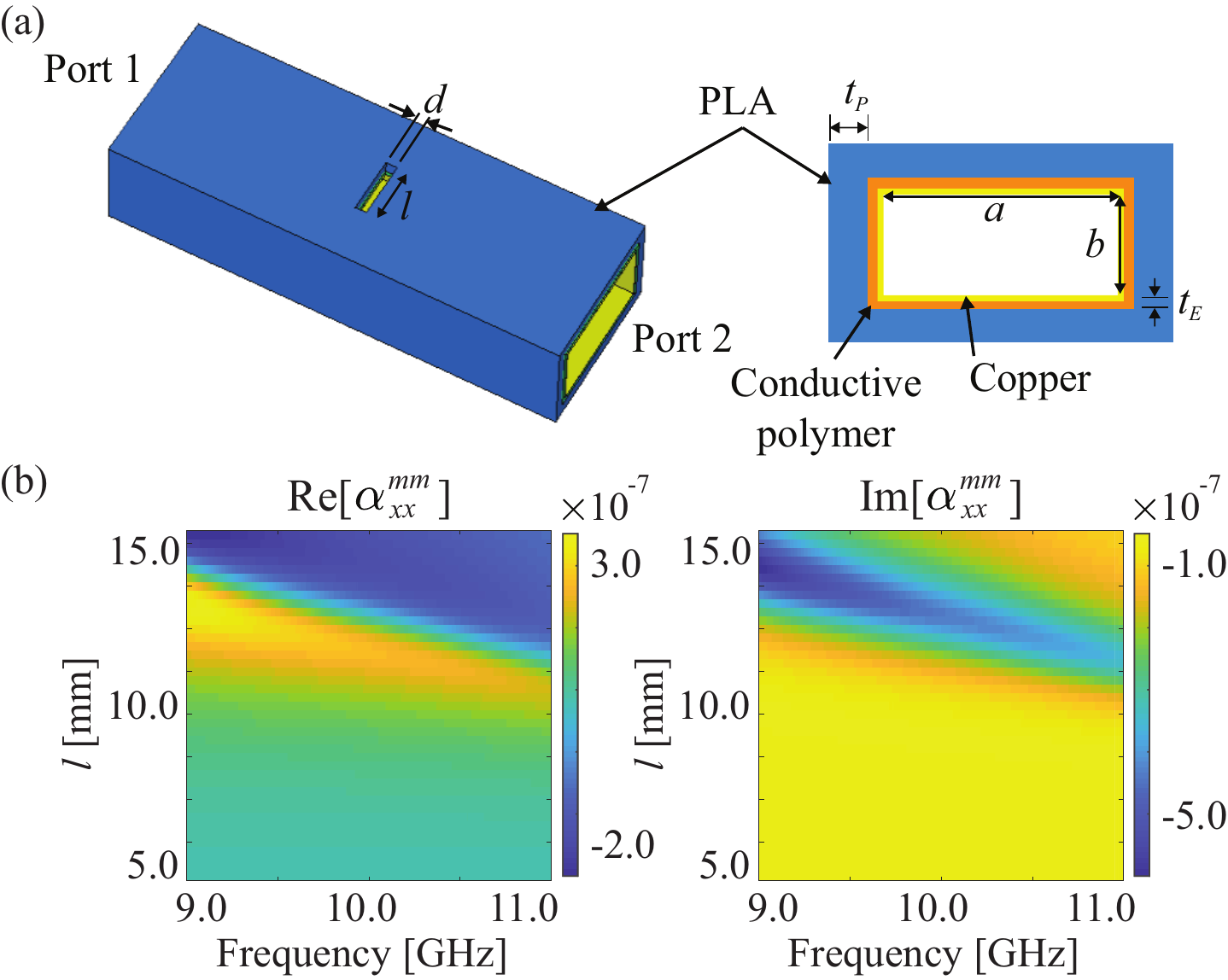}}
\caption{(a) Simulation setup for the polarizability extraction and its side view (inset). A rectilinear slot inserted into top wall of a 3D printed rectangular waveguide is characterized using the effective polarizability, i.e., $\alpha_{}^{m}$. The 3D printed waveguide structure consists of a PLA layer (blue), conductive polymer layer (orange), and copper layer (yellow). Key design parameters are: $a=22.86$ mm, $b=10.16$ mm, $d=1.0$ mm, $t_{P}=2.0$ mm, $t_{E}=0.8$ mm. The slot length is varied from $5.0$ to $15.0$ mm. (b) The real and imaginary part of the extracted polarizability as functions of slot length and frequency.}
\label{Fig1_Polarizability}
\end{figure}

As an initial step for the design of holographic beamforming antennas, we consider a 3D printed, hollow rectangular waveguide with a rectilinear slot shown in Fig. \ref{Fig1_Polarizability}(a). As shown in Fig. \ref{Fig1_Polarizability}(a), the 3D printed waveguide structure is constructed with three layers of materials; a dielectric layer of polylactic acid (PLA) serving as a supporting material, a conductive polymer material printed on the inner surface of the PLA, and a 0.02-mm-thick copper layer deposited on the inner surface of the conductive polymer. The thickness of the copper layer (i.e., $t_{c} = 0.02$ mm) was chosen such that $t_{c}>2\delta_{s}$ where $\delta_{s}$ represents the skin depth at the operational frequencies (near $10.0$ GHz). This particular stack-up of the layers is selected to ease the metal plating process, which will be discussed further below. Also, the material stack-up allows the construction of hollow waveguide structures that suppress unwanted dielectric losses and excitation of higher-order modes. The inner dimensions of the waveguide are chosen such that the standard WR-90 waveguide adapters can be used to excite the structure. Note that the thickness of the PLA layer is chosen to avoid its warping (or bending) after fabrication, i.e., $t_{P}=2.0$ mm. We initially used a $0.4$-mm-thick conductive polymer layer during the design, while the thickness needed to be increased (thus, $0.8$ mm) to avoid any possible defects in the fabricated antennas. The effects of the increased conductive polymer layer thickness on the fabricated antennas' performance will be discussed later in this section.


The waveguide of the antenna is perforated by an array of rectilinear slots inserted into the upper conducting wall of the waveguide\textemdash including the outer PLA, inner conductive polymer, and copper layers. For the design, a single slot is first characterized by simulation, as shown in Fig. \ref{Fig1_Polarizability}(a). The slot couples energy from the guided mode to free space as radiation. We conceptually equate the slot to an effective polarizable point magnetic dipole with its principal axis oriented along the slot \cite{balanis2005antenna}, having an effective magnetic polarizability of $\alpha_{}^{m}$ \cite{pulido2017polarizability}. Note that the finite thickness of the waveguide wall forming the slot may reduce its coupling to the guided modes \cite{collin1960field}; such effects are fully encapsulated in the effective polarizability description.

The polarizability can be retrieved from the structure shown in Fig. \ref{Fig1_Polarizability}(a) using a full-wave electromagnetic solver (here, CST Microwave Studio is used). In the full-wave analysis, PLA is modeled as a dielectric ($\epsilon_r=2.75$ and $\tan\delta=0.005$), and the conductive polymer material modeled using the conductivity of $\sigma_{E} = 1.67\times 10^{4}$ S/m \cite{yurduseven2017computational}. The waveguide is excited using waveguide ports at both ends of the structure and the scattering (S-) parameters calculated. The polarizability can then be found as  \cite{pulido2017polarizability}
\begin{equation} \label{polarizability}
\begin{aligned}
\alpha_{}^{m} = \frac{jab}{2\beta_{}}\left(S_{21}-S_{11}-1\right),
\end{aligned}
\end{equation}
where $a$, $b$ are the width and height of the waveguide, respectively. $\beta$ represents the propagation constant in the waveguide. During the retrieval process, the width of the slot is fixed to be $d=1.0$ mm, while the slot length $l$ is varied from $5.0$ mm to $15.0$ mm. The width and length of the slot are chosen such that the extracted polarizability (i.e., $\alpha_{}^{m}$) exhibits a Lorentzian resonance \cite{smith2017analysis} near $10.0$ GHz, as shown in Fig. \ref{Fig1_Polarizability}(b). Note that the reference plane at which the S-parameters are measured is set to the center of the slot \cite{pulido2017polarizability}.


The next step of the design is to determine a set of polarizabilities that will realize the desired distribution of effective currents (i.e., magnetic dipole moments) over the aperture, given the antenna's configuration (e.g., the number of slots and their spacing). The holographic design method achieves this goal \cite{smith2017analysis}, allowing straightforward solutions for the required polarizabilities. The needed polarizability of $i$th slot in an antenna can be expressed as,
\begin{equation} \label{ideal_pol}
\begin{aligned}
\alpha_{i,\textrm{req}}^{m} = e^{j\beta x_{i}} e^{jk_{0} \sin\phi_{\textrm{tar}}},
\end{aligned}
\end{equation}
where $k_{0}$ represents the free space wavenumber. $x_i$ is the location of $i$th slot, and $\phi_{\textrm{tar}}$ is the target direction of a directive beam. Note that the required polarizabilities in (\ref{ideal_pol}) cannot be implemented using the Lorentzian-constrained polarizabilities (i.e., $\alpha_{}^{m}$) shown in Fig. \ref{Fig1_Polarizability}(b) due to the inherent coupling of magnitude and phase responses \cite{smith2017analysis,bowen2022optimizing}. Thus, a simple method that maps the required and Lorentzian-constrained polarizabilities is needed. We here employ a mapping technique that finds the polarizabilities by minimizing the Euclidean distance between the required and accessible polarizabilities \cite{bowen2022optimizing}. It should be noted that the holographic design method presented in \cite{smith2017analysis} assumes an infinitesimal thin conducting aperture. Therefore, the holographic design approach does not consider the effects of the dielectric layer (i.e. PLA) near the radiating slots. We find that finite thickness slots often lead to increased sidelobes compared with the analytical predictions using the coupled dipole method without considering the effects of finite thickness of the slots \cite{pulido2017discrete}. Therefore, the obtained slot lengths may need slight adjustments using a numerical optimization to reduce sidelobes, which is considered in this work.

Following the outlined design process, we designed and fabricated three holographic beamforming antennas, with target angles of $\phi_{tar,1}=90^{\circ}$ (i.e., broadside), $\phi_{tar,2}=75^{\circ}$, and $\phi_{tar,3}=70^{\circ}$. To demonstrate the applicability of the design process, we used different sets of the number of radiating slots, their spacing for the designs, and operating frequency, while keeping the total aperture size of $\sim 200$ mm. For the antenna generating a beam at $\phi_{tar,1}=90^{\circ}$, $20$ slots were used with $p_{1}=10.0$ mm spacing (i.e., $0.33\lambda_{0}$ at the operating frequency of $10.0$ GHz). For the antenna with the target angle of $\phi_{tar,2}=75^{\circ}$, $25$ slots were used with $p_{2}=7.5$ mm spacing (i.e., $0.25\lambda_{0}$ at the operating frequency of $10.0$ GHz). For the antenna with the target angle of $\phi_{tar,3}=70^{\circ}$, $30$ slots were used with $p_{3}=6.5$ mm spacing (i.e., $0.23\lambda_{0}$ at the operating frequency of $10.5$ GHz).


\begin{figure}[t]
\centerline{\includegraphics[width=3.6in]{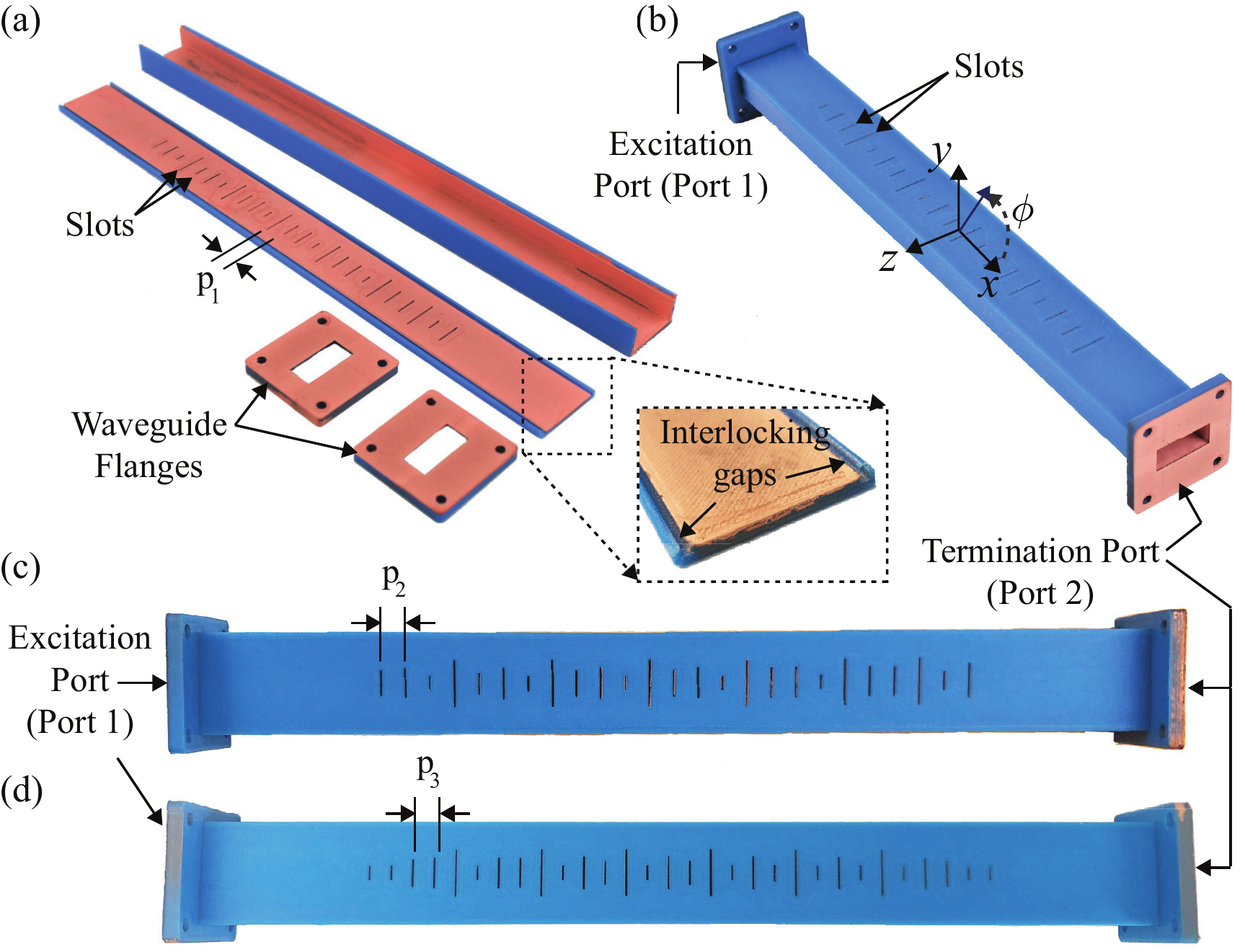}}
\caption{3D printed holographic beamforming antenna with a target direction of $\phi_{1}=90^{\circ}$ (a) before and (b) after assembly. Design parameters are $a=22.86$ mm, $b=10.16$ mm, $d=1.0$ mm, $p_{1}=10.0$ mm. The antenna consists of $20$ slots, and its length of the antenna is $300$ mm. The fabricated antenna is divided into separate parts for the ease in its fabrication and assembled after the fabrication. In the top plate, interlocking gaps were inserted for mechanical bonding of the plate and the waveguide body (inset). At both ends, two flanges are inserted to lock the waveguide structure and connect with the standard WR-90 waveguide-to-coax transitions. (c), (d) Top view of the assembled beamforming antennas with a target direction of $\phi_{2}=75^{\circ}$ and $\phi_{3}=70^{\circ}$. The spacing between adjacent slots in each antenna is $p_{2}=7.5$ mm and $p_{3}=6.5$ mm. The number of slots in the antennas are $25$ and $30$, respectively.}
\label{Fig2_Antenna}
\end{figure}

The fabricated antennas are shown in Figs. \ref{Fig2_Antenna}(a)-(d), with their printing process indicated in Fig. \ref{Fig2_Antenna}(a). As shown in Fig. \ref{Fig2_Antenna}(a), the designed antennas were divided into four separate parts including a top plate with the slots, the main body of the waveguide, and two flanges, for ease of fabrication. During the fabrication, the conductive polymer material, loaded in one of the two extrusion heads, was used to construct the inner surface layer within the slotted waveguides. Meanwhile, PLA, loaded in the other extrusion head, was used to construct the outer shell of the prototype. The dual extrusion capability allows printing of the PLA and Electrifi components in a single process. All four parts were printed without the need of support structure or post processing and easily assembled into a complete structure. Subsequently, the conductive layer was used as a seed layer to deposit a thin layer of copper on its surface to enhance the conductivity and reduce surface roughness \cite{flowers20173d,kim2019one}. After the plating was complete, the printed objects were rinsed with an aqueous solution containing citric acid. The presence of citric acid prevented the plated copper surface from turning dark during the drying process.

It should be noted that no surface activation is needed prior to electroplating as the Electrifi layer is conductive. Also, the copper is plated only on the surface of the Electrifi layer while the PLA is not coated with copper. In this manner, a selective plating was achieved without masking, which may be used to plate more complex 3D structures. After fabrication, the antennas were assembled mechanically using interlocking gaps inserted into the top plates (see inset of Fig. \ref{Fig2_Antenna}(a)). We find that such interlocking gaps help prevent unwanted warping of the 3D printed structures, given their thickness (i.e., $t_{P}=2.0$ mm). The assembled antennas with the target direction of $90^{\circ}$, $75^{\circ}$, and $70^{\circ}$ are respectively depicted in Figs. \ref{Fig2_Antenna}(b)-(d).

The performance of the 3D printed antennas depends on the fabrication process, in addition to the structure. Improvements of the print parameters (e.g., print temperature, platform temperature, extrusion width, infill percentage, print speed, layer height) were thus pursued to achieve higher surface conductivity and lower surface roughness. Table \ref{Table1} summarizes the major printing parameters for the antenna fabrication. Also, we note that the thickness of the conductive and dielectric layers, and their assembly into a complete prototype were refined to ensure a high quality print and electrical properties, while maintaining good mechanical strength.


\begin{table}[b!]
\centering
\caption{The optimized 3D printing parameters for fabrication of the beamforming antennas shown in Figs. \ref{Fig2_Antenna}(a)-(c).} \label{Table1} \centering
\begin{tabular}{l|l}
  \toprule[1pt]
  3D Print Parameters & Values \\
  \midrule[1pt]
   Print Temperature    & Electrifi: 140 $^{\circ}$C \\
                        & PLA: 210 $^{\circ}$C \\ \hline
   Platform Temperature & 45 $^{\circ}$C \\ \hline
   Print Speed          & Electrifi: 15 mm/s \\
                        & PLA: 30 mm/s \\ 
   \bottomrule[1pt]
\end{tabular}
\end{table}

The far-field patterns of the fabricated antennas can be obtained by taking the fast Fourier transform (FFT) of the near-fields measured over a plane proximate to the antenna aperture \cite{yaghjian1986overview}. Near-field data was measured using an NSI planar near-field scanner with an open-ended probe antenna. During the measurement, both ends of the fabricated antenna were connected with standard WR-90 waveguide-to-coaxial adapters. Excitation port (i.e., port 1) was used to feed each of the antennas, with the other port (i.e., port 2) terminated using a matched load. For comparison, we simulated the designed antennas using the full-wave electromagnetic solver, CST Microwave Studio. For the simulations, a $0.4$-mm-thick Electrifi layer was assumed with its inner surface modeled as copper with conductivity $\sigma_{} = 5.8\times 10^{7}$ S/m. Also, we simulated the fabricated antennas using a $0.8$-mm-thick Electrifi layer and deposited copper layer, modeled with conductivity of $\sigma_{c} = 2.5\times 10^{5}$ S/m and surface roughness ($10$ $\mu$m RMS) \cite{kim2019one}. In the simulations, we used a ``waveguide port'' at both ends of the waveguide to excite the structure.


\begin{figure}[t!]
\centerline{\includegraphics[width=2.8in]{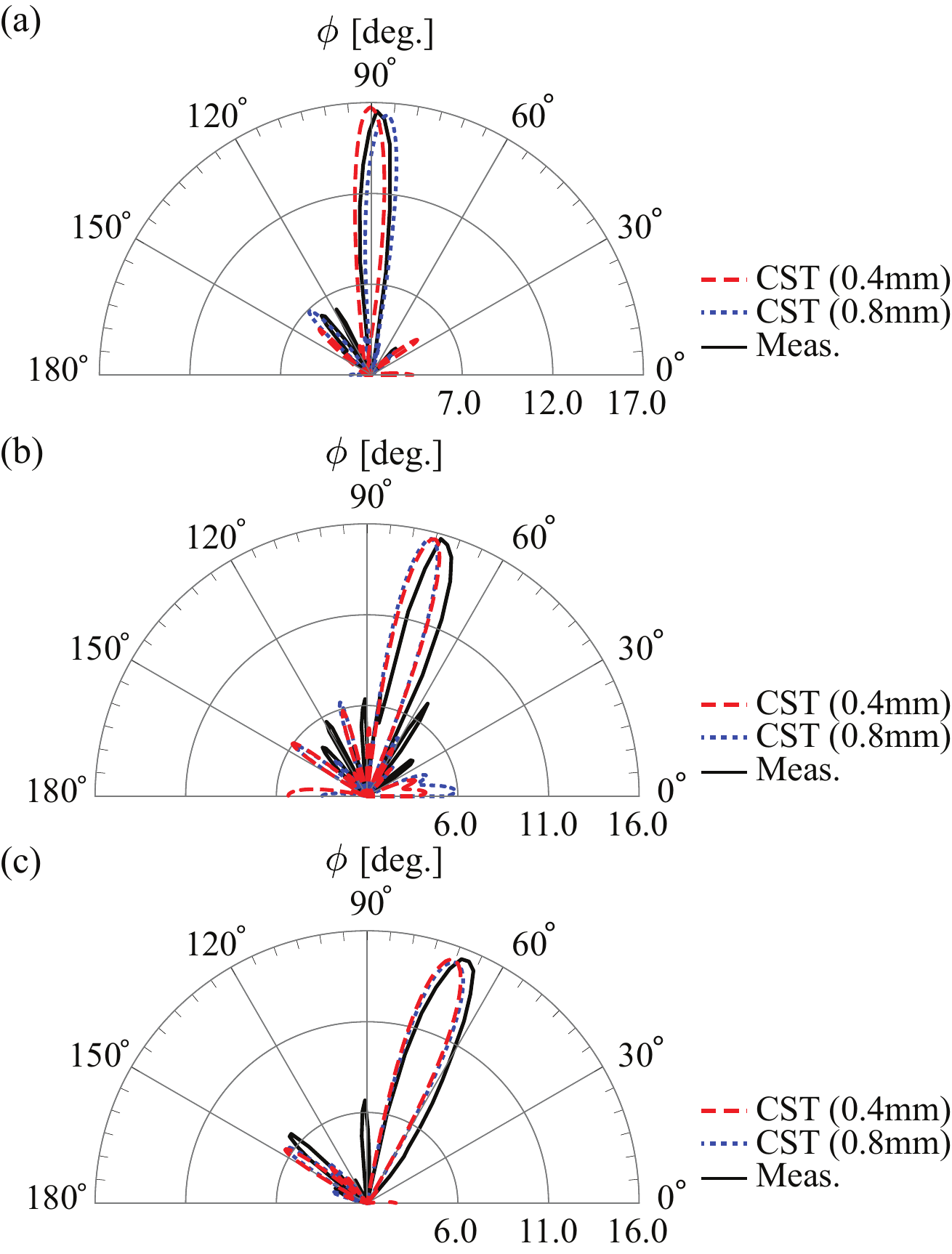}}
\caption{Measured and simulated directivity patterns of the beamforming antenna with a target direction of (a) $\phi_{tar,1}=90^{\circ}$ at $10.0$ GHz, (b) $\phi_{tar,2}=75^{\circ}$ at $10.0$ GHz, and (c) $\phi_{tar,3}=70^{\circ}$ at $10.5$ GHz. In the full-wave simulations, two different Electrifi layer thickness (i.e., $0.4$ and $0.8$ mm) was used.}
\label{Fig3_Patterns}
\end{figure}

Figures \ref{Fig3_Patterns}(a)-(c) compare the simulated and measured directivity patterns of the fabricated antennas at the operating frequencies, demonstrating quite good agreement. As depicted in Fig. \ref{Fig3_Patterns}(a), for the antenna designed to generate a beam at $\phi_{tar,1}=90^{\circ}$, the simulated ($0.4$ and $0.8$ mm-thick Electrifi) and measured peak directivities were $16.7$ dBi at $\phi=89.0^{\circ}$, $16.2$ dBi at $\phi=86.1^{\circ}$, and $16.6$ dBi at $\phi=88.8^{\circ}$, respectively. The sidelobe levels of the simulated and measured patterns were $-11.2$ dB, $-9.9$ dB and $-10.3$ dB, respectively. For the antenna designed to generate a beam at $\phi_{tar,2}=75^{\circ}$, the simulated ($0.4$ and $0.8$ mm-thick Electrifi) and measured peak directivities were $15.8$ dBi at $\phi=74.9^{\circ}$, $15.6$ dBi at $\phi=75.9^{\circ}$, and $15.7$ dBi at $\phi=74.0^{\circ}$, respectively, as shown in Fig. \ref{Fig3_Patterns}(b). The sidelobe levels of the simulated and measured patterns were measured to be $-10.5$ dB, $-9.4$ dB, and $-8.3$ dB, respectively. For the antenna designed to generate a beam at $\phi_{tar,3}=70^{\circ}$, the simulated ($0.4$ and $0.8$ mm-thick Electrifi) and measured peak directivities were $15.2$ dBi at $\phi=70.2^{\circ}$, $15.1$ dBi at $\phi=69.9^{\circ}$, and $15.4$ dBi at $\phi=68.9^{\circ}$, respectively, as shown in Fig. \ref{Fig3_Patterns}(b). The sidelobe levels of the simulated and measured patterns were measured to be $-8.8$ dB, $-8.7$ dB, and $-8.8$ dB, respectively. Note that the designed antennas with the thinner Electrifi layer (i.e., $0.4$ mm) generate beams at the target directions with small offsets. However, increased sidelobes and decreased beam pointing accuracy were observed for the thicker Electrifi layer (i.e., $0.8$ mm).

\begin{figure}[t!]
\centerline{\includegraphics[width=2.9in]{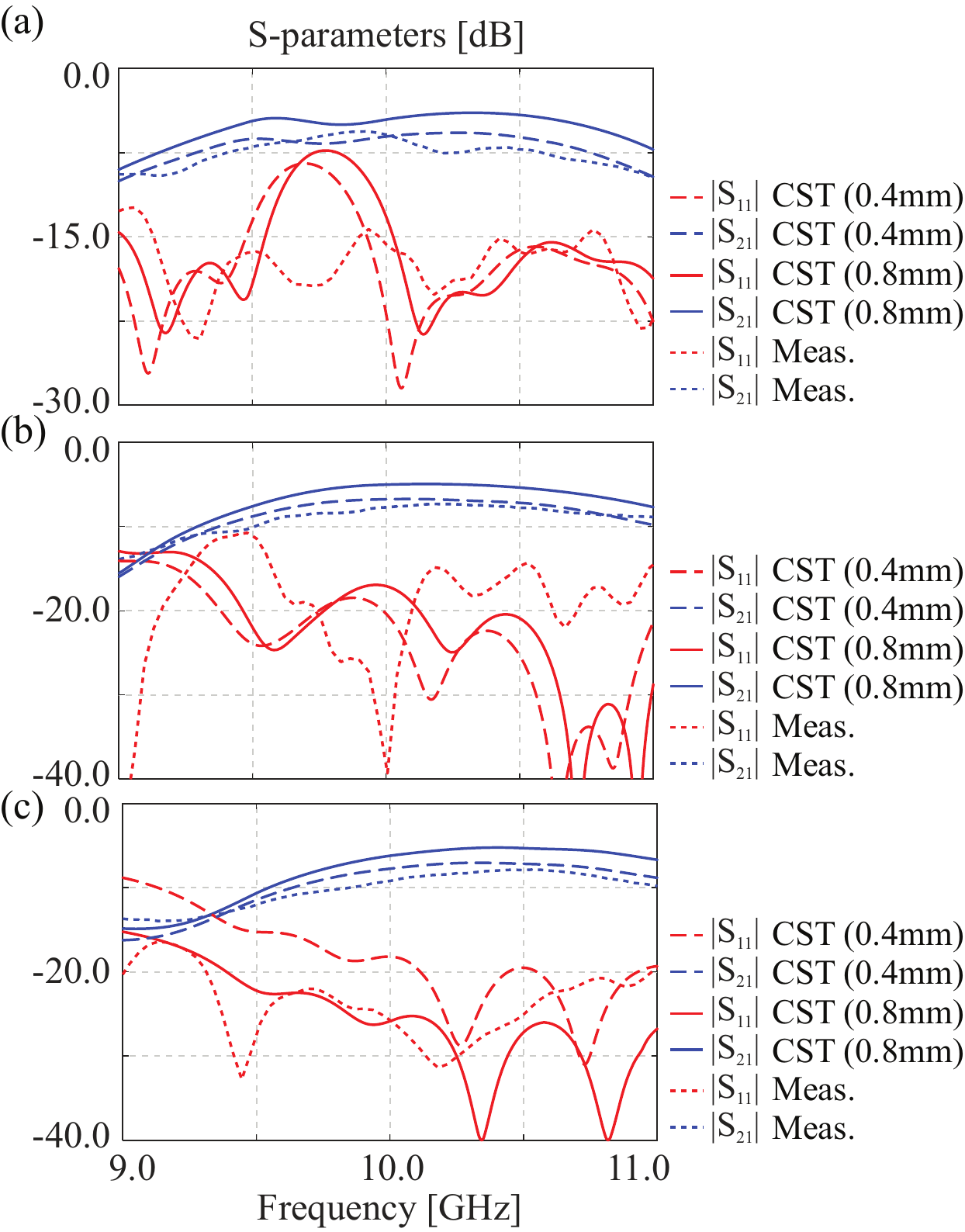}}
\caption{The measured and simulated S-parameters of the desgiend antenna generating a beam at (a) $\phi_{tar,1}=90^{\circ}$, (b) $\phi_{tar,2}=75^{\circ}$, and (c) $\phi_{tar,3}=70^{\circ}$. Two different Electrifi layer thickness (i.e., $0.4$ and $0.8$ mm) was used in the full-wave analyses.}
\label{Fig4_Sparameters}
\end{figure}

The simulated and measured S-parameters of the designed antennas are shown in Figs. \ref{Fig4_Sparameters}(a)-(c). As shown in Figs. \ref{Fig4_Sparameters}(a)-(c), a reasonably good level of impedance matching (i.e., $|S_{11}|<-10$ dB) is observed for the antennas near the operating frequency. Also, $|S_{21}|$ remains low for the fabricated and simulated antennas, indicating that only the portion of the excitation power was dissipated in the matched load. Mismatches between the simulated and measured S-parameters can be reduced by incorporating a measured RMS value of the surface roughness of the electroplated polymer layer into the full-wave analyses. The measurement of the surface roughness can be performed using an optical profiler \cite{kim2019one}, a task that remains as an important step for our future work.


\section{Conclusion}
In this paper, we have demonstrated the design and fabrication of 3D printed, holographic beamforming antennas operating near $10$ GHz. The antennas were designed by using rectilinear slots inserted into the top wall of a hollow rectangular waveguide made of an outer PLA dielectric and an inner layer of electroplated conductive polymer. We designed three beamforming antennas following the holographic beamforming principles and verified their operation. The design examples show that a hybrid process of 3D printing and electroplating can provide a path toward designing high-performance beamforming antennas. The proposed antennas and fabrication method can find applications in wireless systems that require rapid prototyping and testing of complex electromagnetic structures.

\section*{Acknowledgment}
This work was supported by a NASA STTR (Contract No. 80NSSC19C0572) and Multi3D.

\bibliographystyle{IEEEtran}
\bibliography{references.bib}

\begin{thebibliography}{10}
\providecommand{\url}[1]{#1}
\csname url@samestyle\endcsname
\providecommand{\newblock}{\relax}
\providecommand{\bibinfo}[2]{#2}
\providecommand{\BIBentrySTDinterwordspacing}{\spaceskip=0pt\relax}
\providecommand{\BIBentryALTinterwordstretchfactor}{4}
\providecommand{\BIBentryALTinterwordspacing}{\spaceskip=\fontdimen2\font plus
\BIBentryALTinterwordstretchfactor\fontdimen3\font minus
  \fontdimen4\font\relax}
\providecommand{\BIBforeignlanguage}[2]{{%
\expandafter\ifx\csname l@#1\endcsname\relax
\typeout{** WARNING: IEEEtran.bst: No hyphenation pattern has been}%
\typeout{** loaded for the language `#1'. Using the pattern for}%
\typeout{** the default language instead.}%
\else
\language=\csname l@#1\endcsname
\fi
#2}}
\providecommand{\BIBdecl}{\relax}
\BIBdecl

\bibitem{oliner1959guided}
A.~Oliner and A.~Hessel, ``Guided waves on sinusoidally-modulated reactance
  surfaces,'' \emph{IRE Trans. Antennas Propag.}, vol.~7, no.~5, pp. 201--208,
  1959.

\bibitem{sievenpiper2005holographic}
D.~Sievenpiper, J.~Colburn, B.~Fong, J.~Ottusch, and J.~Visher, ``Holographic
  artificial impedance surfaces for conformal antennas,'' in \emph{Antennas and
  Propagation Society International Symposium, 2005 IEEE}, vol.~1.\hskip 1em
  plus 0.5em minus 0.4em\relax IEEE, 2005, pp. 256--259.

\bibitem{fong2010scalar}
B.~H. Fong, J.~S. Colburn, J.~J. Ottusch, J.~L. Visher, and D.~F. Sievenpiper,
  ``Scalar and tensor holographic artificial impedance surfaces,'' \emph{IEEE
  Trans. Antennas and Propag.}, vol.~58, no.~10, pp. 3212--3221, 2010.

\bibitem{Minatti2011}
G.~Minatti, F.~Caminita, M.~Casaletti, and S.~Maci, ``Spiral leaky-wave
  antennas based on modulated surface impedance,'' \emph{IEEE Trans. Antennas
  Propag.}, vol.~59, no.~12, pp. 4436--4444, 2011.

\bibitem{patel2011printed}
A.~M. Patel and A.~Grbic, ``A printed leaky-wave antenna based on a
  sinusoidally-modulated reactance surface,'' \emph{IEEE Trans. Ant. Propag.},
  vol.~59, no.~6, pp. 2087--2096, 2011.

\bibitem{gonzalez2017multibeam}
D.~Gonz{\'a}lez-Ovejero, G.~Minatti, G.~Chattopadhyay, and S.~Maci, ``Multibeam
  by metasurface antennas,'' \emph{IEEE Transactions on Antennas and
  Propagation}, vol.~65, no.~6, pp. 2923--2930, 2017.

\bibitem{yurduseven2017dual}
O.~Yurduseven and D.~R. Smith, ``Dual-polarization printed holographic
  multibeam metasurface antenna,'' \emph{IEEE Ant. Wireless Propag. Lett.},
  vol.~16, pp. 2738--2741, 2017.

\bibitem{smith2017analysis}
D.~R. Smith, O.~Yurduseven, L.~P. Mancera, P.~Bowen, and N.~B. Kundtz,
  ``Analysis of a waveguide-fed metasurface antenna,'' \emph{Phys. Rev. Appl.},
  vol.~8, no.~5, p. 054048, 2017.

\bibitem{boyarsky2021electronically}
M.~Boyarsky, T.~Sleasman, M.~F. Imani, J.~N. Gollub, and D.~R. Smith,
  ``Electronically steered metasurface antenna,'' \emph{Sci. Rep.}, vol.~11,
  no.~1, pp. 1--10, 2021.

\bibitem{stevenson2016metamaterial}
R.~Stevenson, M.~Sazegar, A.~Bily, M.~Johnson, and N.~Kundtz, ``Metamaterial
  surface antenna technology: Commercialization through diffractive
  metamaterials and liquid crystal display manufacturing,'' in \emph{Proc. Int.
  Congr. Adv. Electromagn. Mater. Microw. Opt. (METAMATERIALS)}, 2016, pp.
  349--351.

\bibitem{stevenson2018high}
R.~A. Stevenson, D.~Fotheringham, T.~Freeman, T.~Noel, T.~Mason, and S.~Shafie,
  ``High-throughput satellite connectivity for the constant contact vehicle,''
  in \emph{Proc. 48th Eur. Microw. Conf. (EuMC)}, 2018, pp. 316--319.

\bibitem{devadithya2017gpu}
S.~Devadithya, A.~Pedross-Engel, C.~M. Watts, N.~I. Landy, T.~Driscoll, and
  M.~S. Reynolds, ``{GPU}-accelerated enhanced resolution 3-{D} sar imaging
  with dynamic metamaterial antennas,'' \emph{IEEE Trans. Microw. Theory
  Techn.}, vol.~65, no.~12, pp. 5096--5103, 2017.

\bibitem{staff2019holographic}
P.~Staff, ``Holographic beam forming and phased arrays,'' \emph{Pivotal
  Commware, Inc., Kirkland, WA, USA}, [Online], Available:
  https://pivotalcommware.com/technology.

\bibitem{johnson2015sidelobe}
M.~C. Johnson, S.~L. Brunton, N.~B. Kundtz, and J.~N. Kutz, ``Sidelobe
  canceling for reconfigurable holographic metamaterial antenna,'' \emph{IEEE
  Trans. Antennas and Propag.}, vol.~63, no.~4, pp. 1881--1886, 2015.

\bibitem{zhang2016design}
S.~Zhang, ``Design and fabrication of 3{D}-printed planar fresnel zone plate
  lens,'' \emph{Electron. Lett.}, vol.~52, no.~10, pp. 833--835, 2016.

\bibitem{wu20193}
G.-B. Wu, Y.-S. Zeng, K.~F. Chan, S.-W. Qu, and C.~H. Chan, ``3-{D} printed
  circularly polarized modified fresnel lens operating at terahertz
  frequencies,'' \emph{IEEE Trans. Ant. Propag.}, vol.~67, no.~7, pp.
  4429--4437, 2019.

\bibitem{tak20183}
J.~Tak, A.~Kantemur, Y.~Sharma, and H.~Xin, ``A 3-{D}-printed {W}-band slotted
  waveguide array antenna optimized using machine learning,'' \emph{IEEE Ant.
  Wireless Propag. Lett.}, vol.~17, no.~11, pp. 2008--2012, 2018.

\bibitem{yurduseven2017computational}
O.~Yurduseven, P.~Flowers, S.~Ye, D.~L. Marks, J.~N. Gollub, T.~Fromenteze,
  B.~J. Wiley, and D.~R. Smith, ``Computational microwave imaging using 3{D}
  printed conductive polymer frequency-diverse metasurface antennas,''
  \emph{IET Microw., Antennas Propag.}, vol.~11, no.~14, pp. 1962--1969, 2017.

\bibitem{le20163d}


\bibitem{jun20183}
S.~Y. Jun, A.~Elibiary, B.~Sanz-Izquierdo, L.~Winchester, D.~Bird, and
  A.~McCleland, ``3-{D} printing of conformal antennas for diversity wrist worn
  applications,'' \emph{IEEE Trans. Compon., Packag. Manuf. Techn.}, vol.~8,
  no.~12, pp. 2227--2235, 2018.

\bibitem{kim2019one}
M.~J. Kim, M.~A. Cruz, S.~Ye, A.~L. Gray, G.~L. Smith, N.~Lazarus, C.~J.
  Walker, H.~H. Sigmarsson, and B.~J. Wiley, ``One-step electrodeposition of
  copper on conductive 3{D} printed objects,'' \emph{Addit. Manuf.}, vol.~27,
  pp. 318--326, 2019.

\bibitem{xie2017microwave}
Y.~Xie, S.~Ye, C.~Reyes, P.~Sithikong, B.-I. Popa, B.~J. Wiley, and S.~A.
  Cummer, ``Microwave metamaterials made by fused deposition 3{D} printing of a
  highly conductive copper-based filament,'' \emph{Appl. Phys. Lett.}, vol.
  110, no.~18, p. 181903, 2017.

\bibitem{yurduseven20193d}
O.~Yurduseven, S.~Ye, T.~Fromenteze, B.~J. Wiley, and D.~R. Smith, ``3{D}
  conductive polymer printed metasurface antenna for fresnel focusing,''
  \emph{Designs}, vol.~3, no.~3, p.~46, 2019.

\bibitem{mckerricher2015lightweight}
G.~McKerricher, A.~Nafe, and A.~Shamim, ``Lightweight 3{D} printed microwave
  waveguides and waveguide slot antenna,'' in \emph{Proc. IEEE Int. Symp.
  Antennas Propag. \& USNC/URSI Nat. Radio Sci. Meet.}, 2015, pp. 1322--1323.

\bibitem{zhao2019fully}
K.~Zhao, J.~A. Ramsey, and N.~Ghalichechian, ``Fully 3-{D}-printed
  frequency-scanning slotted waveguide array with wideband power-divider,''
  \emph{IEEE Ant. Wireless Propag. Lett.}, vol.~18, no.~12, pp. 2756--2760,
  2019.

\bibitem{pulido2017polarizability}
L.~Pulido-Mancera, P.~T. Bowen, M.~F. Imani, N.~Kundtz, and D.~Smith,
  ``Polarizability extraction of complementary metamaterial elements in
  waveguides for aperture modeling,'' \emph{Phys. Rev. B}, vol.~96, no.~23, p.
  235402, 2017.

\bibitem{elliot2006antenna}
R.~S. Elliot, \emph{Antenna theory and design}.\hskip 1em plus 0.5em minus
  0.4em\relax John Wiley \& Sons, 2006.

\bibitem{balanis2005antenna}
C.~A. Balanis, \emph{Antenna Theory: Analysis and Design}.\hskip 1em plus 0.5em
  minus 0.4em\relax Wiley, 2005.

\bibitem{collin1960field}
R.~E. Collin, \emph{Field Theory of Guided Waves}.\hskip 1em plus 0.5em minus
  0.4em\relax McGraw-Hill, 1960.

\bibitem{bowen2022optimizing}
P.~T. Bowen, M.~Boyarsky, L.~M. Pulido-Mancera, D.~R. Smith, O.~Yurduseven, and
  M.~Sazegar, ``Optimizing polarizability distributions for metasurface
  apertures with lorentzian-constrained radiators,'' \emph{arXiv preprint
  arXiv:2205.02747}, 2022.

\bibitem{pulido2017discrete}
L.~Pulido-Mancera, M.~F. Imani, and D.~R. Smith, ``Discrete dipole
  approximation for simulation of unusually tapered leaky wave antennas,'' in
  \emph{IEEE MTT-S Int. Microw. Symp. Dig.}, 2017, pp. 409--412.

\bibitem{flowers20173d}
P.~F. Flowers, C.~Reyes, S.~Ye, M.~J. Kim, and B.~J. Wiley, ``3{D} printing
  electronic components and circuits with conductive thermoplastic filament,''
  \emph{Addit. Manuf.}, vol.~18, pp. 156--163, 2017.

\bibitem{yaghjian1986overview}
A.~Yaghjian, ``An overview of near-field antenna measurements,'' \emph{IEEE
  Trans. Ant. Propag.}, vol.~34, no.~1, pp. 30--45, 1986.

\end{thebibliography}

\end{document}